\documentclass{elsart1p}

\usepackage{graphicx}
\usepackage{amssymb}

\begin{document}
\vspace*{-45mm}

\begin{frontmatter}
\title{Fundamental measurements with muons - \\ View from PSI}
\author{Bernhard Lauss \ead{bernhard.lauss@psi.ch} }
\address{
Department of Particles and Matter - UCN Group, \hfill \\
Paul Scherrer Institut, CH-5232 Villigen-PSI, Switzerland
}

\begin{abstract}
Muons can serve as probes to precisely determine fundamental parameters
of the Standard Model or search for `new physics'. The high intensity muon beams
at the Paul Scherrer Institut (PSI) allow for precision 
measurements and searches for rare or forbidden processes. 
Both types of experiments challenge
the Standard Model in a way complementary to high energy physics.
We give a short overview of recent results and ongoing experiments at PSI, 
and of ideas for the future.
\end{abstract}

\end{frontmatter}
%

%%%%%%%%%%%%%%%%%%%%%%%%%%%%%%%%%%%%%%%%%%%%%%%%%%%%%%%%%%%%%%%%%%%%%%%%%%%%%%%
% main text

\section{Introduction}

In 2008 the Paul Scherrer Institut (PSI)\footnote{www.psi.ch} 
celebrated its 20$^{th}$ anniversary and
many years of delivering high intensity muon beams. 
Several upgrades made the 
590 MeV/51 MHz ring cyclotron 
to be up to today the most powerful proton accelerator
of its kind in the world,
which delivers several 10$^8$ muons per second to experiments.
The accelerator runs now routinely with 2.0 mA proton current and 
was already pushed to 2.15 mA for tests.
In the near future running at 2.3 mA is foreseen, and an extensive program 
was launched to boost the operating proton current to 2.6 mA, by 2011, 
and ultimately to 3.0 mA, 
envisaged for 2012 \cite{wagner-nim}. 
Precision experiments should benefit
from a correspondingly increased muon intensity.

\section{Search for the decay  $\mu^+ \rightarrow e^+ + \gamma$}

Charged lepton-flavor conservation has been empirically
verified to a high precision, but is not a consequence of a known underlying symmetry.
The decay $\mu \rightarrow e \gamma$ is lepton-flavor violating
and hence,
excluding neutrino flavor mixing, 
forbidden within the Standard Model (SM).
Neutrino masses and mixing, which is established now, 
introduce a contribution to this decay within the SM,
however, on an unmeasurably small level of 
order $\sim$10$^{-55}$ \cite{meg-proposal}. 
On the other hand, there are several attractive theories beyond the SM, 
such as supersymmetry,
which generally predict lepton-flavor-violating processes at a level within
today's experimental reach. 
A corresponding experimental signal would be free of SM background 
and hence a clear indication for `new physics'.

\begin{figure}
\vspace*{-5mm}
\begin{center}
{\includegraphics[angle=0,scale=0.6]{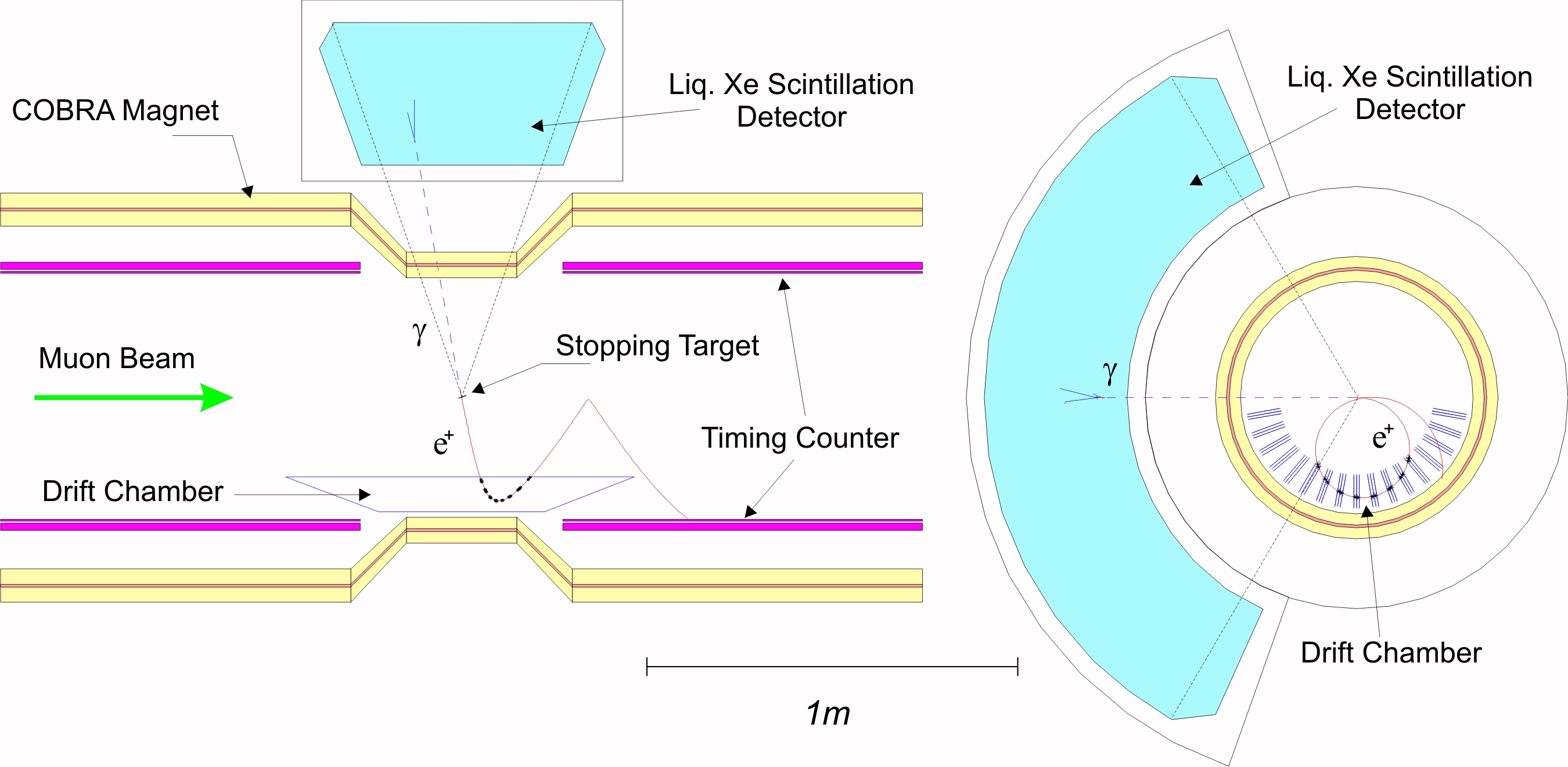}}
\caption{\slshape
Sketch of the MEG experiment's main components \cite{meg-proposal}.}
\label{meg-apparatus}
\end{center}
\end{figure}

The goal of the MEG experiment at PSI \cite{meg-proposal} is to reach a 
sensitivity of 10$^{-13}$, 
improving the present limit \cite{Brooks1999} by almost 2 orders of magnitude.
Consequently one needs a detector managing a challenging 
high muon stop rate up to $10^8$ muons/s. 
The experimental principle is based
on the simultaneous detection of the back-to-back emitted
mono-energetic decay positron and gamma.
The positrons are detected in high rate drift-chambers located 
in a magnetic field
for momentum determination and in scintillation counters for timing.
The gammas are detected in the world's largest liquid xenon scintillation counter,
as sketched in Fig.\ref{meg-apparatus}.
Excellent timing, energy and spatial resolution 
for both reaction products are required to beat the main background
caused by ordinary muon decay and pile-up.
2008 saw the first months of physics run of MEG and the 
accumulated statistics looks promising to already
improve the present limit on $\mu \rightarrow e \gamma$ significantly.

%%%%%%%%%%%%%%%%%%%%%%%%%%%%%%%%%%%%%%%%%%%%%%%%%%%%%%%%%%%%%%%%%%%%%%%%%%%%%%%%%

\section{Muon lifetime measurements}

The Fermi constant $G_F$ describes the strength of the
charged-current weak interaction. 
Along with the fine structure constant $\alpha$ 
and the $Z$-boson mass, it is one of the three pillars of the electroweak 
Standard Model and
directly related to the electroweak gauge coupling \cite{marciano}.
The most precise determination of $G_F$
is based on the mean lifetime of the positive muon, $\tau_{\mu}$, and
can be extracted from:
\begin{equation}
\hspace*{40mm} \frac{1}{\tau_{\mu}} = \frac {G_F^2 m_{\mu}^5}{192 \pi^3}(1 + \Delta q)
\end{equation}
with $\Delta q$ representing higher order QED and hadronic corrections as well as
finite-lepton-mass phase space factors, which have
only recently been computed to a sub-ppm level \cite{stuart-rittbergen}.
A first computation of order $\alpha^2$ using a finite electron mass 
shifted the value of $\Delta q$ by another 0.43ppm \cite{czarnecki}.
Hence, a comparably precise experimental determination of $\tau_{\mu}$ 
is highly desirable.

\begin{figure}
\vspace*{-5mm}
{\includegraphics[angle=0,scale=0.5]{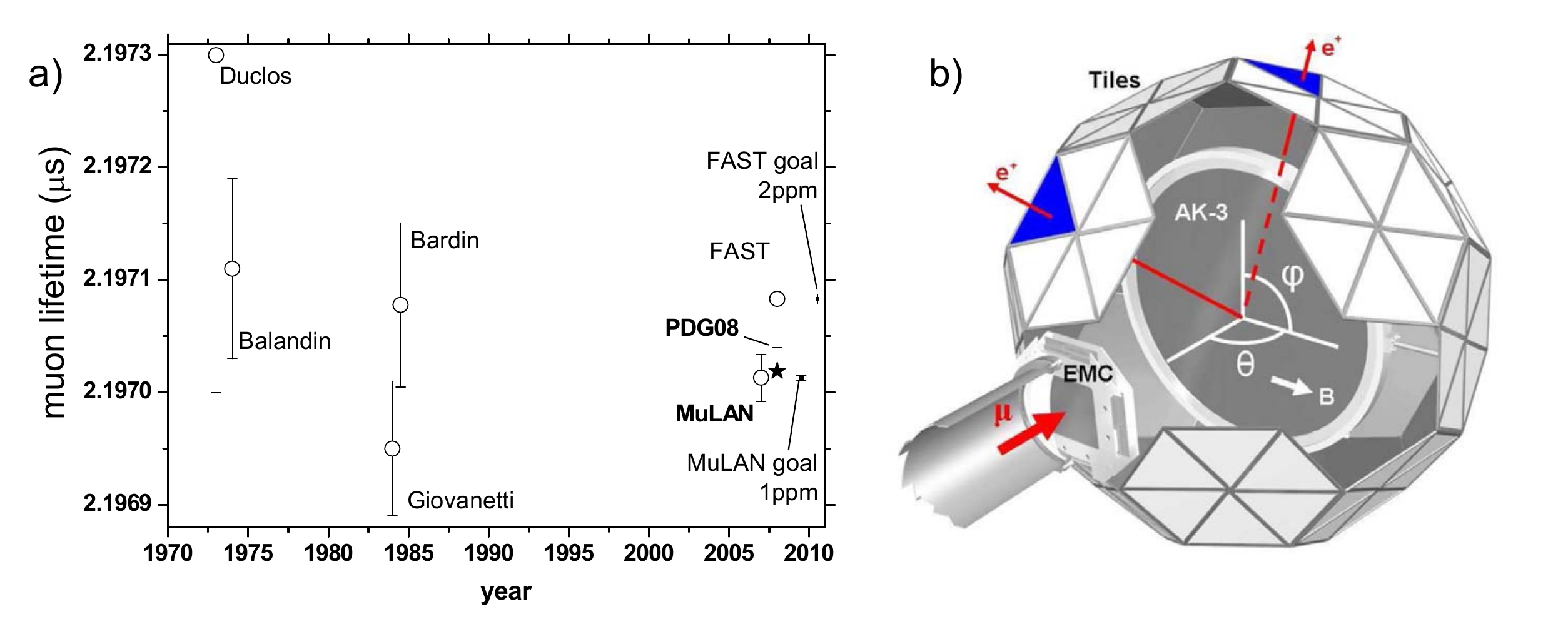}}
\caption{\slshape
a) The improvement of the muon lifetime over the last 40 years 
\cite{mulan-prl, fast-plb, mu-life},
together with the 2008 PDG average (star) and the goals of
the PSI experiments, with errors hardly visible on this scale.
A 1ppm error on $\tau_{\mu}$ translates into a 
0.5ppm error on $G_F$.\newline
b) Sketch of the MuLan detector 
(EMC - entrance muon counter, AK-3 - the target disc, 
tiles - the parts of the scintillation counter) \cite{mulan-prl}.
}
\label{mu-lifetime-plot}
\end{figure}

The MuLan experiment \cite{mulan-prl} installed 
a muon beam kicker 
on the PiE3 beamline at PSI,
which allows after directing positive muons onto 
a target for a selectable time period (e.g. 5 $\mu$s), 
to steer away the beam for the following, for instance, 22 $\mu$s,
The decay positrons are recorded in a soccer-ball shaped detector 
(see Fig.\ref{mu-lifetime-plot}b) made from 
170 double-layer scintillator tiles, 
which are read out via custom-made 500 MHz FADC modules
able to separate pulse pile-up events on the ns level. 
Systematic issues, caused by positron detection differences in the counters, 
due to polarized muons precessing in the earth's magnetic field, are dealt with 
via measurements in different targets, 
which are in a homogeneous magnetic field and either
fully maintain the muon polarization (silver), 
depolarize the muons to a large extent (sulphur), 
or cause a very fast muon precession due to an internal 
few Tesla high magnetic field 
(Arnokrome$^{TM}$-III\footnote{a proprietary Fe-Cr-Co alloy 
of Arnold Engineering Co., Marengo, IL, USA}).
Several 10$^{12}$ muon decays were recorded for each target.
The first MuLan result, based on part of the data 
has set a new precision benchmark, as shown 
in Fig.\ref{mu-lifetime-plot}a.
Additionally, several dedicated systematic measurements are presently under analysis.
The final precision goal on $\tau_{\mu}$ is 1ppm, 
which translates into a 0.5ppm precision on $G_F$.

The FAST experiment \cite{fast-plb} relies on the detection of the full decay sequence
$\pi \rightarrow \mu \rightarrow e$ and corresponding times
in a fast imaging target made of 32 x 48 pixels, constructed
from plastic scintillator bars in a homogeneous B field. 
This approach allows a good control of muon
polarization effects. FAST is scheduled to achieve a statistics
of several $10^{11}$ in 2008/2009. Its goal is a 2ppm measurement of $\tau_{\mu}$.
As a by-product, FAST can also measure the $\pi^+$ lifetime and improve the
present world average.

%%%%%%%%%%%%%%%%%%%%%%%%%%%%%%%%%%%%%%%%%%%%%%%%%%%%%%%%%%%%%%%%%%%%%%%%%%%%%%%%%%%%%%%%%

\section{Muon capture measurements}

\begin{figure}
\vspace*{-5mm}
\begin{center}
{\includegraphics[angle=0,scale=0.53]{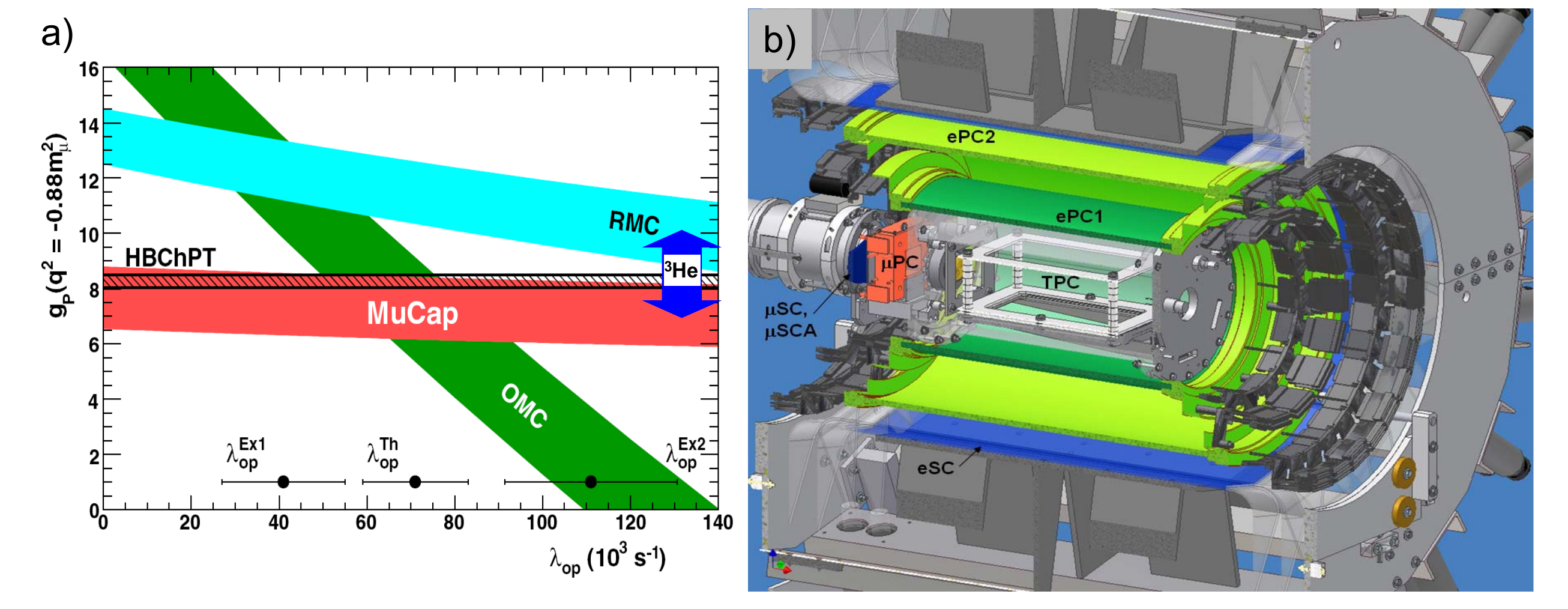}}
\caption{\slshape
a) Present knowledge on the proton's pseudoscalar coupling constant. Explanantion see text.\newline
b) The MuCap detector: $\mu$SC/$\mu$PC - muon entrance counters, TPC - time projection chamber = target, 
ePC, eSC - electron tracking and timing counters.}
\label{gp-plot}
\end{center}
\end{figure}

The determination of the proton's weak pseudoscalar coupling constant 
$g_P$ has been the driving force behind decades of muon capture measurements. 
The PSI result on the muon capture rate in $^3$He \cite{ackerbauer} 
has set a precision landmark in this field.
However, with 3 involved nucleons some questions still remained
in the precise theoretical prediction.
A specially exciting turn came with the precise TRIUMF results from a measurement of
radiative muon capture (RMC) in hydrogen \cite{rmc-triumf},
which disagreed with theory and results derived from ordinary muon capture (OMC)
measurements \cite{omc-saclay}, as shown in Fig.\ref{gp-plot}a.
After decades of worldwide experimental efforts, 
the MuCap experiment has achieved the first unambiguous determination of
the proton's pseudoscalar coupling $g_P$ \cite{mucap-prl} and 
has solved a longstanding discrepancy.
The result is in excellent agreement with
recent calculations based on 
heavy baryon chiral perturbation theory (HBChPT) \cite{meissner}. 
Experimental determinations of $g_P$ depend on
the ortho-para transition rate $\lambda_{op}$ in the $p\mu p$ molecule. 
The most precise previous measurement of ordinary muon capture (OMC) \cite{omc-saclay}
and the RMC experiment \cite{rmc-triumf} both depend significantly on
the value of $\lambda_{op}$, 
which itself is poorly known due to mutually
inconsistent experimental \cite{lopex} and theoretical \cite{lopth} results. 
In contrast, the first MuCap result for $g_P$\cite{mucap-prl} 
is almost independent of molecular effects.

The MuCap result was only possible with an enduring joint effort and a rigorous
experimental technique \cite{exa}. The setup is shown in Fig.\ref{gp-plot}b.
The active target, a time projection chamber, was filled 
with 10 bar of ultra-pure (high $Z$ contamination in the few ppb range) 
and isotopically pure hydrogen \cite{chups-nim}.
Muon stops and corresponding decay electron tracks
were recorded in 3 dimensions, 
which allowed for very selective cuts and
hence an unprecedented control and possible study of systematic effects.
Specifically, muon capture events on high $Z$ elements are even on the 
ppb contamination level visible in the MuCap detector. 
Target conditions were selected in order to control effects due to 
muonic molecule formation and muon catalyzed fusion. 
The muon capture rate was finally determined by comparing the 
lifetime of negative muons in hydrogen with the positive muon
lifetime from the more precise result \cite{mulan-prl}.
In order to extract $g_P$, the singlet muon capture rate 
was compared to two recent 
calculations \cite{mucapture-calculation} 
adding the newly calculated radiative correction \cite{marciano-radiative}.
Presently, the full data set of more than 10$^{10}$ recorded $\mu^-$ stops in hydrogen 
is being analyzed in a blind analysis,
with the final precision goal of 1$\%$ on the singlet muon capture rate
in hydrogen.

\begin{figure}
\vspace*{-5mm}
\begin{center}
{\includegraphics[angle=0,scale=0.52]{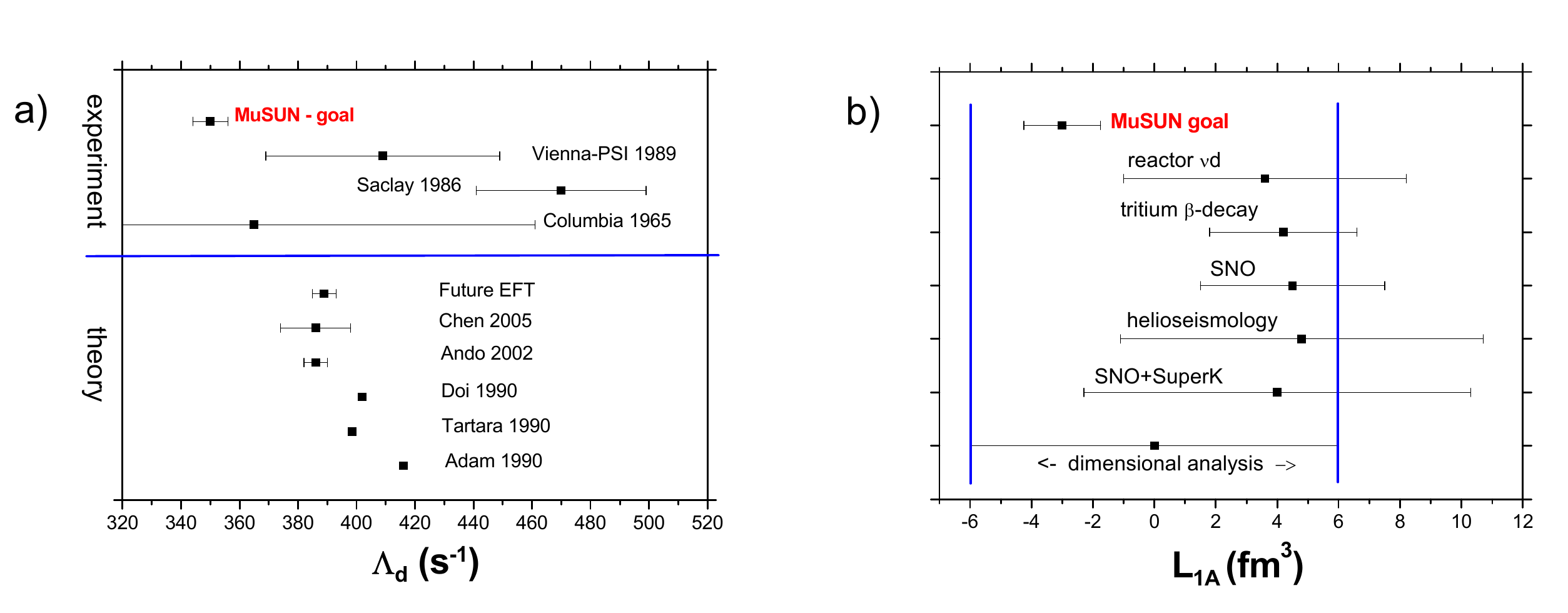}}
\caption{\slshape
a) Experimental determinations of
the doublet muon capture rate on the deuteron
in comparison with recent calculations. 
Meson exchange currents contribute $\sim$20 s$^{-1}$ to the total rate.
b) Estimations on the axial two-body current term $L_{1A}$. All methods up to 
now include certain assumptions and approximations which might be questioned.
All references are given in \cite{musun-proposal}
}
\label{l1a-plot}
\end{center}
\end{figure}

Knowing $g_P$ from MuCap facilitates the interpretation of the 
doublet muon capture rate on the deuteron ($\Lambda_d$), 
the measurement goal of the
recently started MuSun experiment \cite{exa, musun-proposal, claude08} 
which also aims at 1$\%$ precision.
Such a result would allow a precise test of modern effective 
field theories and would represent the most stringent
test of electro-weak interaction in a two-body system \cite{ando}. 
Moreover, it would allow determination of the axial two-body
current term which scales with the low energy constant
denoted $L_{1A}$ (or \^d$^R$) \cite{ando}.
This parameter is of astrophysical interest, as it appears
in the same form in the cross-section for 
i) muon capture on the deuteron, 
ii) $pp$ fusion -- the main fusion reaction in the sun--, 
and iii) neutrino - deuteron reactions, which are the 
detection reactions used by the Sudbury Neutrino Observatory \cite{sno}.
Hence, via the absolute neutrino rates, 
the precise $\mu$d capture rate determination 
would `calibrate' the sun.
Existing experimental results on $\Lambda_d$ are not 
precise enough (Fig.\ref{l1a-plot}a)
and also using other sources leaves
the present experimental knowledge on $L_{1A}$ rather sparse, 
as shown in Fig.\ref{l1a-plot}b.
With precision neutrino physics on the horizon the precise knowledge
of $L_{1A}$ will be necessary, as it also influences 
the determination of $\delta m^2_{21}$ and $\Theta_{13}$\cite{balantekin}.

The experimental principle of MuSun will follow the successful MuCap 
approach, but for control and optimization of muonic molecule formation
and muon catalyzed fusion reactions in deuterium 
one has to use a high density cryo-target
at $\sim$30K \cite{musun-proposal, claude08}. 
$dd$ fusion reactions occur at much higher rates than muon capture
and hence represent a severe background, but also allow 
to monitor the hyperfine populations of the muonic atom \cite{petitjean-dd}. 
High $Z$ target purity will be even more critical as in MuCap. 
In a first engineering run a new pad-based TPC was successfully tested
with high purity deuterium in late 2008.

%%%%%%%%%%%%%%%%%%%%%%%%%%%%%%%%%%%%%%%%%%%%%%%%%%%%%%%%%%%%%%%%%%%%%%%%

\section{Further fundamental measurements and ideas}

There are several other precision measurements of fundamental parameters
ongoing or under discussion at PSI, which either will test the Standard Model 
or search for new physics.

\begin{itemize}

\item{
The muonic Lamb-shift experiment \cite{pohl} is preparing for its
first physics data taking in 2009, and wants to 
precisely determine the proton charge radius
via observation of the 2p-2s energy difference in muonic hydrogen.
}

\item{
A precise test of the electron-muon universality
is being performed within the PEN experiment 
and corresponding data are being analyzed \cite{dinko}.
}

\item{
The search for the lepton-flavor-violating process $\mu \rightarrow eee$ would be 
a sensitive search for new physics and complement the present MEG activity. 
There is an ongoing discussion
on 2 suggested experimental approaches,
how to obtain a sensitivity which improves the 
present experimental limit by roughly 3 orders of magnitude \cite{vanderschaaf-chipp}.
}

\item{
A sensitive search for a CP violating muon electric dipole moment (EDM) was suggested
in \cite{kirch-edm} using a compact storage ring,
which could make use of PSI's high muon intensity and reach a sensitivity 
of 5$\times$10$^{-23}$ e$\cdot$cm.
In this way it would test` new physics' 
and pave the way for higher sensitivity tests of muon
and other charged particle EDMs.
}

\item{
The discussion about dark matter and dark energy has also put interest in particles
decaying into mirror worlds, other dimensions or to other branes.
Hence the decay products would be invisible.
A search for the invisible decay of muons was suggested in \cite{invisible-decay}, and might be also
searched for by using the MuCap setup.
}

\item{High brightness muon beams would also allow a first test of the gravitational interaction
of antimatter of a purely leptonic system, which involves second generation particles, 
namely muonium ($\mu^+ e^-$) \cite{kirch-gravity}.
}

\end{itemize}

\vspace*{5mm}

Given these and more ideas, one can be sure that
precision measurements using muons,
at PSI and other facilities in the world,
will also in future contribute 
to a deeper understanding and testing of the Standard Model and
provide a fair chance to first find `new physics' beyond our
presently accepted theory.

\section{Acknowledgements}

I would like to express my sincere gratitude for advice, 
support and many stimulating discussions to 
K. Kirch, P. Kammel, D. Hertzog, S. Ritt,
C. Casella, C.~Petitjean, M. Seidel, 
and especially all members of the MuCap, MuLan and MuSun Collaborations.

%%%%%%%%%%%%%%%%%%%%%%%%%%%%%%%%%%%%%%%%%%%%%%%%%%%%%%%%%%%%%%%%%%%%%%%%%%

%

%

\begin{thebibliography}{00}
%
\bibitem{wagner-nim}
W.~Wagner et al., Nucl. Instr. and Meth. A (2008) in press.
\hfill

\bibitem{meg-proposal}
L.M.~Barkov et al., 
Search for $\mu^+ \rightarrow e^+ \gamma$ down to 10$^{-14}$ branching ratio,
The MEG experiment, PSI proposal R-99-05, 1999;
G.~Signorelli, these proceedings.
\hfill


\bibitem{Brooks1999}
M.L.~Brooks et al., Phys. Rev. Lett. 83 (1999) 1521.
\hfill


\bibitem{marciano}
W.J.~Marciano, Phys.Rev.D 60 (1999) 093006.
\hfill



\bibitem{stuart-rittbergen}
T.~van Ritbergen, R.G.~Stuart, Nucl. Phys. B 564 (2000) 343;
T.~van Ritbergen, R.G.~Stuart, Phys. Lett. B 437 (1998) 201.
\hfill


\bibitem{czarnecki}
A.~Pak, A.~Czarnecki, Phys. Rev. Lett. 100 (2008) 241807.
\hfill




\bibitem{mulan-prl}
D.B.~Chitwood, et al. (MuLan Collaboration), Phys. Rev. Lett. 99 (2007) 032001.
\hfill



\bibitem{fast-plb}
A.~Barczyk et al., Phys. Lett.B 663 (2008) 172.
\hfill


\bibitem{mu-life}
J.~Duclos et al., Phys. Lett. B47 (1973) 491;
M.P.~Balandin et al., Sov. Phys. JETP 40 (1974) 811;
G.~Bardin et al., Phys. Lett. B137 (1984) 135; 
K.~Giovanetti et al., Phys. Rev. D29,(1984) 343.
\hfill


\bibitem{ackerbauer}
P.~Ackerbauer et al., Phys. Lett. B417 (1998) 224.
\hfill



\bibitem{rmc-triumf}
D.H.~Wright et al., Phys. Rev. C 57 (1998) 373.
\hfill


\bibitem{mucap-prl}
A.~Andreev et al. (MuCap Collaboration), Phys. Rev. Lett. 99 (2007) 032002.
\hfill


\bibitem{meissner}
V.~Bernard, L.~Elouadrhiri, and U.-G.~Meissner, J. Phys. G 28 (2002) R1.
\hfill

\bibitem{omc-saclay}
G.~Bardin et al., Nucl. Phys. A352 (1981) 365.
\hfill


\bibitem{lopex}
G.~Bardin et al., Phys. Lett. B 104, 320 (1981) 320;
J.H.D.~Clark et al., Phys. Rev. Lett. 96 (2006) 073401.
\hfill


\bibitem{lopth}
D.D.~Bakalov et al., Nucl. Phys. A384, 302 (1982) 302.
\hfill


\bibitem{exa}
P.~Kammel in: 
Proceedings of the
EXA'02 - International Workshop on Exotic Atoms - Future Perspectives, Vienna, Austria, 2002;
Austrian Academy of Sciences Press, Vienna, 2003; arXiv:nucl-ex/0304019; 
B.~Lauss in:
Proceedings of the EXA'05 - Intern. Conference on Exotic Atoms, 
Vienna, Austria, 2005, Austrian Academy of Sciences Press, Vienna 2005; arXiv:nucl-ex/0401005.
\hfill


\bibitem{chups-nim}
V.A.~Ganzha et al., Nucl. Instr. and Meth. A578 (2007) 485,
\hfill


\bibitem{mucapture-calculation}
V.~Bernard, T.R.~Hemmert, U.-G.~Meissner, Nucl. Phys. A686, (2001) 290;
S.~Ando, F.~Myhrer, K.~Kubodera, Phys. Rev. C63 (2000) 015203.
\hfill


\bibitem{marciano-radiative}
A.~Czarnecki, W.J.~Marciano, A.~Sirlin Phys. Rev. Lett. 99 (2007) 032003.
\hfill


\bibitem{ando}
S.~Ando et al., Phys. Lett. B 533 (2002) 25; 
J.W. Chen et al., Phys. Rev. C 72 (2005) 061001.
\hfill


\bibitem{musun-proposal}
V.A.~Andreev et al., Muon Capture in deuterium, the MuSun experiment,
PSI Proposal R-08-01, 2008.
\hfill

\bibitem{claude08}
C.~Petitjean in: Proceedings of the International Symposium on Pulsed Neutron and
Muon Sciences, Mito, Japan 2008; Nucl. Inst. and Meth. A 2008, in press.
\hfill

\bibitem{petitjean-dd}
C.~Petitjean et al., Hyperfine Int. 118 (1999) 127;
B.~Lauss et al., Hyperfine Int. 118 (1999) 79.
\hfill


\bibitem{sno}
B.~Aharmim et al., Phys. Rev. C75 (2007) 045502;
J.-W.~Chen, K.M.~Heeger, R.G.~Hamish Robertson, Phys.Rev. C67 (2003) 025801.
\hfill


\bibitem{balantekin}
A.B.~Balantekin, H.~Yuksel, 
Phys. Rev. C68 (2003) 055801 and
Int.J. Mod. Phys., E14 (2005) 39.
\hfill


\bibitem{pohl}
R.~Pohl et al., AIP Conf. Proc. 796 (2005) 253.
\hfill


\bibitem{dinko}
D.~Po\^cani\'c, these proceedings.
\hfill

\bibitem{vanderschaaf-chipp}
A.~van der Schaaf, CHIPP Meeting, Lausanne, Switzerland, 2008.
\hfill

\bibitem{kirch-edm}
A.~Adelmann et al., arxiv:hep-ex/0606034.
\hfill

\bibitem{invisible-decay}
S.~Gninenko, Phys. Rev. D76 (2007) 055004.
\hfill

\bibitem{kirch-gravity}
K.~Kirch, arxiv:hep-ex/0702143.
\hfill



%
\end{thebibliography}
\end{document}